\documentstyle[12pt]{article}
\begin{document}

\def\ds{\displaystyle}
\def\beq{\begin{equation}}
\def\eeq{\end{equation}}
\def\bea{\begin{eqnarray}}
\def\eea{\end{eqnarray}}
\def\beeq{\begin{eqnarray}}
\def\eeeq{\end{eqnarray}}
\def\ve{\vert}
\def\vel{\left|}
\def\ver{\right|}
\def\nnb{\nonumber}
\def\ga{\left(}
\def\dr{\right)}
\def\aga{\left\{}
\def\adr{\right\}}
\def\lla{\left<}
\def\rra{\right>}
\def\rar{\rightarrow}
\def\nnb{\nonumber}
\def\la{\langle}
\def\ra{\rangle}
\def\ba{\begin{array}}
\def\ea{\end{array}}
\def\tr{\mbox{Tr}}
\def\ssp{{\Sigma^{*+}}}
\def\sso{{\Sigma^{*0}}}
\def\ssm{{\Sigma^{*-}}}
\def\xis0{{\Xi^{*0}}}
\def\xism{{\Xi^{*-}}}
\def\qs{\la \bar s s \ra}
\def\qu{\la \bar u u \ra}
\def\qd{\la \bar d d \ra}
\def\qq{\la \bar q q \ra}
\def\gGgG{\la g^2 G^2 \ra}
\def\q{\gamma_5 \not\!q}
\def\x{\gamma_5 \not\!x}
\def\g5{\gamma_5}
\def\sb{S_Q^{cf}}
\def\sd{S_d^{be}}
\def\su{S_u^{ad}}
\def\ss{S_s^{??}}
\def\sbp{{S}_Q^{'cf}}
\def\sdp{{S}_d^{'be}}
\def\sup{{S}_u^{'ad}}
\def\ssp{{S}_s^{'??}}
\def\sig{\sigma_{\mu \nu} \gamma_5 p^\mu q^\nu}
\def\fo{f_0(\frac{s_0}{M^2})}
\def\ffi{f_1(\frac{s_0}{M^2})}
\def\fii{f_2(\frac{s_0}{M^2})}
\def\O{{\cal O}}
\def\sl{{\Sigma^0 \Lambda}}
\def\es{\!\!\! &=& \!\!\!}
\def\ar{&+& \!\!\!}
\def\ek{&-& \!\!\!}
\def\cp{&\times& \!\!\!}
\def\se{\!\!\! &\simeq& \!\!\!}
\def\kpm{&\pm& \!\!\!}
\def\kmp{&\mp& \!\!\!}


\def\simlt{\stackrel{<}{{}_\sim}}
\def\simgt{\stackrel{>}{{}_\sim}}


\title{
         {\Large
                 {\bf
Model independent analysis of $\Lambda$ baryon polarizations in
$\Lambda_b \rar \Lambda \ell^+ \ell^-$ decay 
                 }
         }
      }

\author{\vspace{1cm}\\
{\small T. M. Aliev$^a$ \thanks
{e-mail: taliev@metu.edu.tr}\,\,,
A. \"{O}zpineci$^b$ \thanks
{e-mail: ozpineci@ictp.trieste.it}\,\,,
M. Savc{\i}$^a$ \thanks
{e-mail: savci@metu.edu.tr}} \\
{\small a Physics Department, Middle East Technical University, 
06531 Ankara, Turkey}\\
{\small b  The Abdus Salam International Center for Theoretical Physics,
I-34100, Trieste, Italy} }
\date{}

\begin{titlepage}
\maketitle
\thispagestyle{empty}

\begin{abstract}
We present the model independent analysis of $\Lambda$ baryon polarizations
in the $\Lambda_b \rar \Lambda \ell^+ \ell^-$ decay. The sensitivity of the
averaged $\Lambda$ polarizations to the new Wilson coefficients is studied.
It is observed that there exist certain regions of the new Wilson
coefficients where the branching ratio coincides with the standard model
prediction, while the $\Lambda$ baryon polarizations deviate from the
standard model results remarkably.
\end{abstract}

~~~PACS numbers: 12.60.--i, 13.30.--a
\end{titlepage}

\section{Introduction}
Flavor--changing neutral current (FCNC) $b \rar s(d) \ell^+ \ell^-$
transitions provide potentially the most sensitive and stringiest test 
for the standard model (SM) in the flavor sector at loop level, since FCNC
transitions are forbidden in the SM in the Born approximation.
At the same time these decays are very sensitive to the
new physics beyond the SM. New physics appear in rare
decays through the Wilson coefficients which can take values different from
their SM counterpart or through the new operator structures in an effective
Hamiltonian \cite{R5201}.

First measurements of the $B \rar X_s \gamma$ decay were reported by CLEO
Collaboration \cite{R5202} and at present more precise measurements are
currently being carried out in the experiments at B factories \cite{R5203}.
Exclusive decay involving the $b \rar s \gamma$ transition has been measured
in \cite{R5204}--\cite{R5206}. After these measurements of the radiative
decay induced by the $b \rar s \gamma$ transition, main interest has been
focused on the rare decays induced by the $b \rar s \ell^+ \ell^-$
transition, which have relatively large branching ratio in the SM. These
decays have been extensively studied in the SM and its various extensions
\cite{R5207}--\cite{R5222}. 

The exclusive $B \rar K^\ast (K) \ell^+ \ell^-$ decays, which are described 
by $b \rar s \ell^+ \ell^-$ transition at inclusive level, have been widely
studied in literature (see \cite{R5222}--\cite{R5226} and references
therein). Recently BaBar Collaboration announced evidence of the 
$B \rar K \ell^+ \ell^-$ and $B \rar K^\ast \ell^+ \ell^-$ decays with the
branching ratios ${\cal B}(B \rar K \ell^+ \ell^-) =
(0.78^{+0.24+0.11}_{-0.20-0.18}) \times 10^{-6}$,
${\cal B}(B \rar K^\ast \ell^+ \ell^-) = 
(1.68^{+0.68}_{-0.58} \pm 0.28) \times 10^{-6}$ \cite{R5227}. The $B \rar K
\ell^+ \ell^-$ decay has been also observed at BELLE detector \cite{R5228}
with the branching ratio ${\cal B}(B \rar K \ell^+ \ell^-) =
(0.75^{+0.25}_{-0.21} \pm 0.09) \times 10^{-6}$. Another exclusive decay
which is described at inclusive level by the $b \rar s \ell^+ \ell^-$
transition is the baryonic $\Lambda_b \rar \Lambda \ell^+ \ell^-$ decay.
Interest to the baryonic decays can be attributed to the fact that, unlike
mesonic decays, they could maintain the helicity structure of the effective
Hamiltonian for the $b \rar s$ transition. Note that, new physics effects
in the $\Lambda_b \rar \Lambda \gamma$ decay were studied in 
\cite{R5229a}.

In this work we analyze the possibility of searching for new physics in the
heavy baryon $\Lambda_b \rar \Lambda \ell^+ \ell^-$ decay by measuring the
polarization of $\Lambda$ baryon, using the most general model independent
form of the effective Hamiltonian. It should be mentioned here that the
sensitivity of the lepton polarization to the new Wilson coefficients, which
are responsible for the existence of new physics beyond the SM in the 
$B \rar K \ell^+ \ell^-$ and $B \rar K^\ast \ell^+ \ell^-$ and
$\Lambda_b \rar \Lambda \ell^+ \ell^-$ decays, is investigated in
\cite{R5229}, \cite{R5230} and \cite{R5231}, respectively, using the most
general form of the effective Hamiltonian. It is shown in these works that
the lepton polarizations are really very sensitive to the new physics
effects. 

The paper is organized as follows. In section
2, using the most general form of the effective Hamiltonian , the general
expressions for the longitudinal and normal polarizations
of the $\Lambda$ baryon are derived. Section 3 is devoted to the study of
the dependence of the  $\Lambda$ polarizations on the new Wilson
coefficients.

\section{Lepton polarizations}

At quark level, the matrix element of the $\Lambda_b \rar \Lambda \ell^+ \ell^-$ 
decay is described by the $b \rar s \ell^+ \ell^-$ transition. The effective
Hamiltonian responsible for the $b \rar s \ell^+ \ell^-$ transition can be
written in terms of twelve model independent four--Fermi interactions as
\cite{R5224,R5232}

\bea
\label{e1}
{\cal M} \es \frac{G \alpha}{\sqrt{2} \pi} V_{tb}V_{ts}^\ast \Bigg\{
C_{SL} \bar s_R i \sigma_{\mu\nu} \frac{q^\nu}{q^2} b_L \bar \ell \gamma^\mu
\ell + C_{BR} \bar s_L i \sigma_{\mu\nu} \frac{q^\nu}{q^2} b_R \bar \ell
\gamma^\mu \ell + C_{LL}^{tot} \bar s_L \gamma_\mu b_L \bar \ell_L
\gamma^\mu \ell_L \nnb \\
\ar C_{LR}^{tot} \bar s_L \gamma_\mu b_L \bar \ell_R  
\gamma^\mu \ell_R + C_{RL} \bar s_R \gamma_\mu b_R \bar \ell_L
\gamma^\mu \ell_L + C_{RR} \bar s_R \gamma_\mu b_R \bar \ell_R
\gamma^\mu \ell_R \nnb \\
\ar C_{LRLR} \bar s_L b_R \bar \ell_L \ell_R +
C_{RLLR} \bar s_R b_L \bar \ell_L \ell_R +
C_{LRRL} \bar s_L b_R \bar \ell_R \ell_L +
C_{RLRL} \bar s_R b_L \bar \ell_R \ell_L \nnb \\
\ar C_T \bar s \sigma_{\mu\nu} b \bar \ell \sigma^{\mu\nu} \ell +
i C_{TE} \epsilon_{\mu\nu\alpha\beta} \bar s \sigma^{\mu\nu} b 
\bar \ell \sigma^{\alpha\beta} \ell \Bigg\}~,
\eea
where the subindices $L$ and $R$ stand for the chiral operators 
$L=(1-\gamma_5)/2$ and $R=(1+\gamma_5)/2$.
The coefficients of the first two terms, $C_{SL}$ and
$C_{BR}$ describe the penguin contributions, which correspond to 
$-2 m_s C_7^{eff}$ and $-2 m_b C_7^{eff}$ in the SM, respectively. 
The next four terms with coefficients
$C_{LL}^{tot},~C_{LR}^{tot},~ C_{RL}$ and $C_{RR}$ in Eq. (\ref{e1})
describe vector type interactions. Two of these coefficients
$C_{LL}^{tot}$ and $C_{LR}^{tot}$ contain SM results in the form
$C_9^{eff}-C_{10}$ and $C_9^{eff}-C_{10}$, respectively.
For this reason
we can write \bea
\label{e2}
C_{LL}^{tot} \es C_9^{eff}- C_{10} + C_{LL}~, \nnb \\
C_{LR}^{tot} \es C_9^{eff}+ C_{10} + C_{LR}~,
\eea
where $C_{LL}$ and $C_{LR}$ describe the contributions of new physics. The
next four terms in Eq. (\ref{e1}) with
coefficients $C_{LRLR},~C_{RLLR},~C_{LRRL}$ and $C_{RLRL}$ represent the
scalar type interactions. The remaining last two terms led by the
coefficients $C_T$ and $C_{TE}$ are the tensor type interactions.

The amplitude of the exclusive $\Lambda_b \rar \Lambda\ell^+ \ell^-$ decay
is obtained by calculating the matrix element of ${\cal H}_{eff}$ for the $b
\rar s \ell^+ \ell^-$ transition between initial and final
baryon states $\lla \Lambda \vel {\cal H}_{eff} \ver \Lambda_b \rra$.
We see from Eq. (\ref{e1}) that for
calculating the $\Lambda_b \rar \Lambda\ell^+ \ell^-$ decay amplitude,
the following matrix elements are needed

\bea
&&\lla \Lambda \vel \bar s \gamma_\mu (1 \mp \gamma_5) b \ver \Lambda_b
\rra~,\nnb \\
&&\lla \Lambda \vel \bar s \sigma_{\mu\nu} (1 \mp \gamma_5) b \ver \Lambda_b
\rra~,\nnb \\
&&\lla \Lambda \vel \bar s (1 \mp \gamma_5) b \ver \Lambda_b \rra~.\nnb
\eea

The relevant matrix elements parametrized in terms of the form factors are 
as follows (see \cite{R5233,R5234})

\bea
\label{e3}
\lla \Lambda \vel \bar s \gamma_\mu b \ver \Lambda_b \rra  
\es \bar u_\Lambda \Big[ f_1 \gamma_\mu + i f_2 \sigma_{\mu\nu} q^\nu + f_3  
q_\mu \Big] u_{\Lambda_b}~,\\
\label{e4}
\lla \Lambda \vel \bar s \gamma_\mu \gamma_5 b \ver \Lambda_b \rra
\es \bar u_\Lambda \Big[ g_1 \gamma_\mu \gamma_5 + i g_2 \sigma_{\mu\nu}
\gamma_5 q^\nu + g_3 q_\mu \gamma_5\Big] u_{\Lambda_b}~, \\
\label{e5}
\lla \Lambda \vel \bar s \sigma_{\mu\nu} b \ver \Lambda_b \rra
\es \bar u_\Lambda \Big[ f_T \sigma_{\mu\nu} - i f_T^V \ga \gamma_\mu q^\nu -
\gamma_\nu q^\mu \dr - i f_T^S \ga P_\mu q^\nu - P_\nu q^\mu \dr \Big]
u_{\Lambda_b}~,\\
\label{e6}
\lla \Lambda \vel \bar s \sigma_{\mu\nu} \gamma_5 b \ver \Lambda_b \rra
\es \bar u_\Lambda \Big[ g_T \sigma_{\mu\nu} - i g_T^V \ga \gamma_\mu q^\nu -
\gamma_\nu q^\mu \dr - i g_T^S \ga P_\mu q^\nu - P_\nu q^\mu \dr \Big]
\gamma_5 u_{\Lambda_b}~,
\eea
where $P = P_{\Lambda_b} + P_\Lambda$ and $q= P_{\Lambda_b} - P_\Lambda$. 

The form factors of the magnetic dipole operators are defined as 
\bea
\label{e7}
\lla \Lambda \vel \bar s i \sigma_{\mu\nu} q^\nu  b \ver \Lambda_b \rra
\es \bar u_\Lambda \Big[ f_1^T \gamma_\mu + i f_2^T \sigma_{\mu\nu} q^\nu
+ f_3^T q_\mu \Big] u_{\Lambda_b}~,\nnb \\
\lla \Lambda \vel \bar s i \sigma_{\mu\nu}\gamma_5  q^\nu  b \ver \Lambda_b \rra
\es \bar u_\Lambda \Big[ g_1^T \gamma_\mu \gamma_5 + i g_2^T \sigma_{\mu\nu}
\gamma_5 q^\nu + g_3^T q_\mu \gamma_5\Big] u_{\Lambda_b}~.
\eea

Note that, using the identity 
\bea
\sigma_{\mu\nu}\gamma_5 = - \frac{i}{2} \epsilon_{\mu\nu\alpha\beta}
\sigma^{\alpha\beta}~,\nnb
\eea
the second expression in Eq. (\ref{e7}) can be written as
\bea
\lla \Lambda \vel \bar s i \sigma_{\mu\nu}\gamma_5  q^\nu  b \ver \Lambda_b \rra
\es \bar u_\Lambda \Big[ f_T i \sigma_{\mu\nu} \gamma_5 q^\nu \Big]
u_{\Lambda_b}~.\nnb
\eea  
Multiplying (\ref{e5}) and (\ref{e6}) by $i q^\nu$ and comparing with
(\ref{e7}), one can easily obtain the following relations between the form
factors
\bea
\label{e8}
f_2^T \es f_T + f_T^S q^2~,\crcr
f_1^T \es \Big[ f_T^V + f_T^S \ga m_{\Lambda_b} + m_\Lambda\dr \Big] 
q^2~ = - \frac{q^2}{m_{\Lambda_b} - m_\Lambda} f_3^T~,\nnb \\
g_2^T \es g_T + g_T^S q^2~,\\
g_1^T \es \Big[ g_T^V - g_T^S \ga m_{\Lambda_b} - m_\Lambda\dr \Big]
q^2 =  \frac{q^2}{m_{\Lambda_b} + m_\Lambda} g_3^T~.\nnb
\eea 

The matrix element of the scalar (pseudoscalar) operators $\bar s b$ and
$\bar s\gamma_5 b$ can be obtained from (\ref{e3}) and (\ref{e4}) by multiplying both
sides to $q^\mu$ and using equation of motion. Neglecting the mass of the
strange quark, we get
\bea
\label{e9}
\lla \Lambda \vel \bar s b \ver \Lambda_b \rra \es \frac{1}{m_b} 
\bar u_\Lambda \Big[ f_1 \ga m_{\Lambda_b} - m_\Lambda \dr + f_3 q^2
\Big] u_{\Lambda_b}~,\\
\label{a8}
\lla \Lambda \vel \bar s \gamma_5 b \ver \Lambda_b \rra \es \frac{1}{m_b} 
\bar u_\Lambda \Big[ g_1 \ga m_{\Lambda_b} + m_\Lambda\dr \gamma_5 - g_3 q^2
\gamma_5 \Big] u_{\Lambda_b}~.
\eea

Using these definitions of the form factors, for the matrix element
of the $\Lambda_b \rar \Lambda\ell^+ \ell^-$ we get \cite{R5234,R5235}

\bea
\label{e10}
\lefteqn{
{\cal M} = \frac{G \alpha}{4 \sqrt{2}\pi} V_{tb}V_{ts}^\ast \Bigg\{
\bar \ell \gamma^\mu \ell \, \bar u_\Lambda \Big[ A_1 \gamma_\mu (1+\gamma_5) +
B_1 \gamma_\mu (1-\gamma_5) }\nnb \\
\ar i \sigma_{\mu\nu} q^\nu \big[ A_2 (1+\gamma_5) + B_2 (1-\gamma_5) \big]
+q_\mu \big[ A_3 (1+\gamma_5) + B_3 (1-\gamma_5) \big]\Big] u_{\Lambda_b}
\nnb \\
\ar \bar \ell \gamma^\mu \gamma_5 \ell \, \bar u_\Lambda \Big[
D_1 \gamma_\mu (1+\gamma_5) + E_1 \gamma_\mu (1-\gamma_5) +
i \sigma_{\mu\nu} q^\nu \big[ D_2 (1+\gamma_5) + E_2 (1-\gamma_5) \big]
\nnb \\
\ar q_\mu \big[ D_3 (1+\gamma_5) + E_3 (1-\gamma_5) \big]\Big] u_{\Lambda_b}+
\bar \ell \ell\, \bar u_\Lambda \big(N_1 + H_1 \gamma_5\big) u_{\Lambda_b}
+\bar \ell \gamma_5 \ell \, \bar u_\Lambda \big(N_2 + H_2 \gamma_5\big) 
u_{\Lambda_b}\nnb \\
\ar 4 C_T \bar \ell \sigma^{\mu\nu}\ell \, \bar u_\Lambda \Big[ f_T 
\sigma_{\mu\nu} - i f_T^V \big( q_\nu \gamma_\mu - q_\mu \gamma_\nu \big) -
i f_T^S \big( P_\mu q_\nu - P_\nu q_\mu \big) \Big] u_{\Lambda_b}\nnb \\
\ar 4 C_{TE} \epsilon^{\mu\nu\alpha\beta} \bar \ell \sigma_{\alpha\beta}
\ell \, i \bar u_\Lambda \Big[ f_T \sigma_{\mu\nu} - 
i f_T^V \big( q_\nu \gamma_\mu - q_\mu \gamma_\nu \big) -
i f_T^S \big( P_\mu q_\nu - P_\nu q_\mu \big) \Big] u_{\Lambda_b}\Bigg\}~,
\eea
where the explicit forms of the functions $A_i,~B_i,~D_i,~E_i,~H_j$ and $N_j$
$(i=1,2,3$ and $j=1,2)$ are as follows \cite{R5234}

\bea
\label{e11}
A_1 \es \frac{1}{q^2}\ga f_1^T-g_1^T \dr C_{SL} + \frac{1}{q^2}\ga
f_1^T+g_1^T \dr C_{BR} + \frac{1}{2}\ga f_1-g_1 \dr \ga C_{LL}^{tot} +
C_{LR}^{tot} \dr \nnb \\
\ar \frac{1}{2}\ga f_1+g_1 \dr \ga C_{RL} + C_{RR} \dr~,\nnb \\
A_2 \es A_1 \ga 1 \rar 2 \dr ~,\nnb \\
A_3 \es A_1 \ga 1 \rar 3 \dr ~,\nnb \\
B_1 \es A_1 \ga g_1 \rar - g_1;~g_1^T \rar - g_1^T \dr ~,\nnb \\
B_2 \es B_1 \ga 1 \rar 2 \dr ~,\nnb \\
B_3 \es B_1 \ga 1 \rar 3 \dr ~,\nnb \\
D_1 \es \frac{1}{2} \ga C_{RR} - C_{RL} \dr \ga f_1+g_1 \dr +
\frac{1}{2} \ga C_{LR}^{tot} - C_{LL}^{tot} \dr \ga f_1-g_1 \dr~,\nnb \\
D_2 \es D_1 \ga 1 \rar 2 \dr ~, \\
D_3 \es D_1 \ga 1 \rar 3 \dr ~,\nnb \\
E_1 \es D_1 \ga g_1 \rar - g_1 \dr ~,\nnb \\
E_2 \es E_1 \ga 1 \rar 2 \dr ~,\nnb \\
E_3 \es E_1 \ga 1 \rar 3 \dr ~,\nnb \\
N_1 \es \frac{1}{m_b} \Big( f_1 \ga m_{\Lambda_b} - m_\Lambda\dr + f_3 q^2
\Big) \Big( C_{LRLR} + C_{RLLR} + C_{LRRL} + C_{RLRL} \Big)~,\nnb \\
N_2 \es N_1 \ga C_{LRRL} \rar - C_{LRRL};~C_{RLRL} \rar - C_{RLRL} \dr~,\nnb \\
H_1 \es \frac{1}{m_b} \Big( g_1 \ga m_{\Lambda_b} + m_\Lambda\dr - g_3 q^2  
\Big) \Big( C_{LRLR} - C_{RLLR} + C_{LRRL} - C_{RLRL} \Big)~,\nnb \\
H_2 \es H_1 \ga C_{LRRL} \rar - C_{LRRL};~C_{RLRL} \rar - C_{RLRL} \dr~.\nnb
\eea

From the expressions of the above-mentioned matrix elements we observe
that $\Lambda_b \rar\Lambda \ell^+\ell^-$ decay is described in terms of  
many form factors. It is shown in \cite{R5236} that when HQET is applied the
number of independent form factors reduces to two ($F_1$ and
$F_2$) irrelevant of the Dirac structure
of the corresponding operators, i.e., 
\bea
\label{e12}
\lla \Lambda(p_\Lambda) \vel \bar s \Gamma b \ver \Lambda(p_{\Lambda_b})
\rra = \bar u_\Lambda \Big[F_1(q^2) + \not\!v F_2(q^2)\Big] \Gamma
u_{\Lambda_b}~,
\eea
where $\Gamma$ is an arbitrary Dirac structure,
$v^\mu=p_{\Lambda_b}^\mu/m_{\Lambda_b}$ is the four--velocity of
$\Lambda_b$, and $q=p_{\Lambda_b}-p_\Lambda$ is the momentum transfer.
Comparing the general form of the form factors given in Eqs.
(\ref{e3})--(\ref{e9}) with (\ref{e12}), one can
easily obtain the following relations among them (see also \cite{R5233})
\bea
\label{e13}
g_1 \es f_1 = f_2^T= g_2^T = F_1 + \sqrt{r} F_2~, \nnb \\
g_2 \es f_2 = g_3 = f_3 = g_T^V = f_T^V = \frac{F_2}{m_{\Lambda_b}}~,\nnb \\
g_T^S \es f_T^S = 0 ~,\nnb \\
g_1^T \es f_1^T = \frac{F_2}{m_{\Lambda_b}} q^2~,\nnb \\
g_3^T \es \frac{F_2}{m_{\Lambda_b}} \ga m_{\Lambda_b} + m_\Lambda \dr~,\nnb \\
f_3^T \es - \frac{F_2}{m_{\Lambda_b}} \ga m_{\Lambda_b} - m_\Lambda \dr~,
\eea
where $r=m_\Lambda^2/m_{\Lambda_b}^2$.

Having obtained the matrix element for the $\Lambda_b \rar\Lambda \ell^+ 
\ell^-$ decay, our next aim is the calculation of $\Lambda$ baryon polarizations
using this matrix element. For this purpose
we write the $\Lambda$ baryon spin four--vector in terms of a unit vector
$\vec{\xi}$ along the $\Lambda$ baryon spin in its rest frame as
\bea
\label{e14} 
s_\mu = \ga \frac{\vec{p}_\Lambda \cdot \vec{\xi}}{m_\Lambda},
\vec{\xi} + \frac{\vec{p}_\Lambda (\vec{p}_\Lambda \cdot
\vec{\xi})}{E_\Lambda+m_\Lambda} \dr ~,
\eea   
and choose the unit vectors along the longitudinal, transversal and normal
components of the $\Lambda$ polarization to be
\bea
\label{e15}
\vec{e}_L = \frac{\vec{p}_\Lambda}{\vel \vec{p}_\Lambda \ver}~, ~~~
\vec{e}_T = \frac{\vec{p}_\ell\times \vec{p}_\Lambda}
{\vel \vec{p}_\ell\times \vec{p}_\Lambda \ver}~,~~~
\vec{e}_N = \vec{e}_T \times \vec{e}_L~,
\eea
respectively, where  $\vec{p}_\ell$ and $\vec{p}_\Lambda$ are the three
momenta of $\ell$ and $\Lambda$, in the center of mass frame of the 
$\ell^+ \ell^-$ system.

The differential decay rate of the $\Lambda_b \rar \Lambda \ell^+ \ell^-$ decay 
for any spin direction $\vec{\xi}$ along the $\Lambda$ baryon
can be written as
\bea
\label{e16}
\frac{d\Gamma(\vec{\xi})}{ds} = \frac{1}{2}
\ga \frac{d\Gamma}{ds}\dr_0
\Bigg[ 1 + \Bigg( P_L \vec{e}_L + P_N
\vec{e}_N + P_T \vec{e}_T \Bigg) \cdot
\vec{\xi} \Bigg]~,
\eea
where $\ga d\Gamma/ds \dr_0$ corresponds to the unpolarized differential
decay rate, $s=q^2/m_{\Lambda_b}^2$ and    
$P_L$, $P_N$ and $P_T$ represent the longitudinal, normal and 
transversal polarizations of the $\Lambda$ baryon, respectively. 
The unpolarized decay width  in Eq. (\ref{e16}) can be written as

\bea
\label{e17}
\ga \frac{d \Gamma}{ds}\dr_0 = \frac{G^2 \alpha^2}{8192 \pi^5}
\vel V_{tb} V_{ts}^\ast \ver^2 \lambda^{1/2}(1,r,s) v 
\Big[{\cal T}_0(s) +\frac{1}{3} {\cal T}_2(s) \Big]~, 
\eea
where 
$\lambda(1,r,s) = 1 + r^2 + s^2 - 2 r - 2 s - 2 rs$
is the triangle function, $r=m_\Lambda^2/m_{\Lambda_b}^2$ and 
$v=\sqrt{1-4m_\ell^2/q^2}$ is the lepton
velocity. The explicit expressions for ${\cal T}_0$ and ${\cal T}_2$ can be 
found in \cite{R5234}.

The polarizations $P_L$, $P_N$ and $P_T$ are defined as:
\bea
\label{e18}
P_i(q^2) = \frac{\ds{\frac{d \Gamma}{ds}
                   (\vec{\xi}=\vec{e}_i) -
                   \frac{d \Gamma}{ds}
                   (\vec{\xi}=-\vec{e}_i)}}
              {\ds{\frac{d \Gamma}{ds}
                   (\vec{\xi}=\vec{e}_i) +
                  \frac{d \Gamma}{ds}
                  (\vec{\xi}=-\vec{e}_i)}}~,
\eea
where $i = L,N,T$. $P_L$ and $P_N$ are $P$--odd, $T$--even, while
$P_T$ is $P$--even, $T$--odd and $CP$--odd. Note that 
transversal polarization of the $\Lambda$ baryon has already been 
studied in \cite{R5235}.

In the massless lepton limit the explicit expressions of the $P_L$ and
$P_N$ for the $\Lambda$ baryon are:
\bea
P_L \es \frac{16 m_{\Lambda_b}^4 \sqrt{\lambda}}
{{\cal T}_0(s) +\frac{1}{3} {\cal T}_2(s)} \Bigg\{
- 4 m_{\Lambda_b} s \, 
\mbox{\rm Re}[A_1^\ast B_2 - A_2^\ast B_1] \nnb \\
\ek s \, 
\Big( v^2 \mbox{\rm Re}[F_1^\ast H_1] + 
\mbox{\rm Re}[F_2^\ast H_2] \Big) \nnb \\
\ek \frac{4}{3} m_{\Lambda_b} s v^2 \,  
\Big( \sqrt{r}\, \mbox{\rm Re}[A_1^\ast A_2 - B_1^\ast B_2]+
3 \mbox{\rm Re}[D_1^\ast E_2 - D_2^\ast E_1] +
\sqrt{r} \, \mbox{\rm Re}[D_1^\ast D_2 - E_1^\ast E_2] \Big) \nnb \\
\ar \frac{1}{3} \Big\{ 
[3 (1-r+s)- v^2 (1-r-s)] (\vel A_1 \ver^2 - \vel B_1\ver^2 + 
\vel D_1 \ver^2 - \vel E_1 \ver^2) \Big\} \nnb \\
\ek \frac{1}{3} m_{\Lambda_b}^2    
s [3 (1-r+s) + v^2 (1-r-s)]
(\vel A_2 \ver^2 - \vel B_2 \ver^2) \nnb \\
\ek \frac{2}{3} m_{\Lambda_b}^2 s v^2 (2 - 2 r + s) \,
(\vel D_2 \ver^2 - \vel E_2 \ver^2) \nnb \\
\ek \frac{256}{3} m_{\Lambda_b}^2 s v^2 [(1+\sqrt{r})^2 -s] \,
\mbox{\rm Re}[C_T^\ast C_{TE}] 
\mbox{\rm Re}[f_T^\ast f_T^S] \nnb \\
\ek \frac{256}{3} m_{\Lambda_b}
s [3-\sqrt{r} (3 - 2 v^2)] \, \mbox{\rm Re}[C_T^\ast C_{TE}] 
\mbox{\rm Re}[f_T^\ast f_T^V] \nnb \\
\ar \frac{256}{3} [3 - 3 r - (1-r-s) v^2] \,
\mbox{\rm Re}[C_T^\ast C_{TE}] \vel f_T \ver^2 \Bigg\}~, 
\\ \nnb \\
\label{a2}
P_N \es \frac{8 \pi m_{\Lambda_b}^4 v \sqrt{s}}
{{\cal T}_0(s) +\frac{1}{3} {\cal T}_2(s)} \Bigg\{
- 2 (1-r+s) \sqrt{r} \,
\mbox{\rm Re}[A_1^\ast D_1 + B_1^\ast E_1] \nnb \\
\ar 4 (1+\sqrt{r}) [(1-\sqrt{r})^2 -s] \, \Big(
\mbox{\rm Re}[(C_T f_T)^\ast H_1] - 
2 \mbox{\rm Re}[(C_{TE} f_T)^\ast H_2] \Big) \nnb \\
\ar 4 (1-\sqrt{r}) [(1+\sqrt{r})^2 -s] \, \Big(
2 \mbox{\rm Re}[(C_{TE} f_T)^\ast F_1] - 
\mbox{\rm Re}[(C_T f_T)^\ast F_2] \Big) \nnb \\
\ar 4 m_{\Lambda_b} s \sqrt{r} \,
\mbox{\rm Re}[A_1^\ast E_2 + A_2^\ast E_1 +B_1^\ast D_2 +
B_2^\ast D_1] \nnb \\
\ek 2 m_{\Lambda_b}^2 s \sqrt{r} (1-r+s) \, 
\mbox{\rm Re}[A_2^\ast D_2 + B_2^\ast E_2^\ast] \nnb \\
\ek 4 m_{\Lambda_b} [(1-\sqrt{r})^2 -s] s \,
\Big( \mbox{\rm Re}[(C_T f_T^V)^\ast H_1] -  
2 \mbox{\rm Re}[(C_{TE} f_T^V)^\ast H_2] \Big) \nnb \\
\ar 2 (1-r-s) \, \Big(
\mbox{\rm Re}[A_1^\ast E_1 + B_1^\ast D_1] + 
m_{\Lambda_b}^2 s \mbox{\rm Re}[A_2^\ast E_2 + B_2^\ast D_2] \Big) \nnb \\
\ek m_{\Lambda_b} [(1-r)^2-s^2] \,      
\mbox{\rm Re}[A_1^\ast D_2 + A_2^\ast D_1 + B_1^\ast E_2 + 
B_2^\ast E_1] \Bigg\},
\eea
respectively.
For the massive lepton case, expressions for the longitudinal and normal 
polarizations are quite lengthy, and for this reason, they are not presented
in the text. The explicit form of these expressions can be found in 
\cite{R5238a}.

\section{Numerical analysis}

In this section we will study the dependence of the lepton polarizations, as
well as combined lepton polarization to the new Wilson coefficients. The
main input parameters in the calculations are the form factors. Since the
literature lacks exact calculations for the form factors of the $\Lambda_b
\rar \Lambda$ transition, we will use the results from QCD sum rules
approach in combination with HQET \cite{R5236,R5237}, which reduces the number of
quite many form factors into two. The $s$ dependence of these form factors
can be represented in the following way
\bea
\label{e19}
F(q^2) = \frac{F(0)}{\ds 1-a_F s + b_F s^2}~, \nnb
\eea
where parameters $F_i(0),~a$ and $b$ are listed in table 1.
\begin{table}[h]    
\renewcommand{\arraystretch}{1.5} 
\addtolength{\arraycolsep}{3pt}  
$$
\begin{array}{|l|ccc|}
\hline
& F(0) & a_F & b_F \\ \hline
F_1 &
\phantom{-}0.462 & -0.0182 & -0.000176 \\
F_2 &
-0.077 & -0.0685 &\phantom{-}0.00146 \\ \hline
\end{array}
$$
\caption{Transition form factors for $\Lambda_b \rar \Lambda \ell^+ \ell^-$
decay in the QCD sum rules method.}
\renewcommand{\arraystretch}{1}
\addtolength{\arraycolsep}{-3pt}
\end{table}

We use the next--to--leading order logarithmic approximation for the resulting
values of the Wilson coefficients $C_9^{eff},~C_7$ and $C_{10}$ in the SM
\cite{R5238,R5239} at the renormalization point $\mu=m_b$. It should be
noted that, in addition to short distance short distance contribution,
$C_9^{eff}$ receives also long distance contributions from the real $\bar c
c$ resonant states of the $J/\psi$ family. The Wilson coefficient
$C_9^{eff}$ is given by
\bea
\label{e20}
C_9^{eff} = C_9(\mu) + Y_{pert} + \frac{3\pi}{\alpha^2} \widetilde{C}^{(0)}
\sum_{V=J/\psi,\psi^\prime,\cdots} \kappa_i 
\frac{\Gamma(V_i \rar \ell^+ \ell^-) m_{V_i}}
     {m_{V_i}^2 - q^2 - i m_{V_i} \Gamma_{V_i}}~,
\eea 
where $C_9(\mu=m_b)=4.214$ in the next to leading logarithmic order (NLL)
\cite{R5238}, $Y_{pert}(q^2/m_b^2)$ arises from the one--loop matrix
elements of the four--quark operators and its explicit expression can be
found in \cite{R5239}, and $\widetilde{C}^{(0)} = 3\widetilde{C}_1 +
\widetilde{C}_2 + 3 \widetilde{C}_3 + \widetilde{C}_4 + 3\widetilde{C}_5 +
\widetilde{C}_6$. The values of these Wilson coefficients in NLL order can
be found in \cite{R5240}. In Eq. (\ref{e20}), $m_{V_i}$ and $\Gamma_{V_i}$
are the masses and widths of the $J/\psi$ family. The fudge factor
$\kappa_i$ for the lowest resonances are chosen as $\kappa_{J/\psi} = 1.65$
and $\kappa_{\psi^\prime}=2.36$ \cite{R5241}, and for higher resonances the
average of $\kappa_{J/\psi}$ and $\kappa_{\psi^\prime}$ is used. In the
present work we neglect the long distance contributions to the $C_9^{eff}$,
i.e., we restrict ourselves by considering only short distance effects.
We will discuss the influence of the long distance effects in one of our
future works. 

It follows from Eq. (\ref{e17}) that in performing the numerical analysis of 
the $\Lambda$ baryon polarizations, the values of the new Wilson coefficients,
which are responsible for the new physics beyond the SM, are needed. In
further numerical analysis we vary all new Wilson coefficients in the range
$-\vel C_{10} \ver \le C_X \le \vel C_{10} \ver$. The experimental bound on
the branching ratio of the $B \rar K^\ast \mu^+ \mu^-$ \cite{R5227,R5228} and
$B \rar \mu^+ \mu^-$ \cite{R5242} decays suggest that this is the right 
order of magnitude for the vector and scalar interaction coefficients.
As has been mentioned in the introduction section, BaBar and BELLE
Collaborations have presented their preliminary results on the branching ratios of
the $B \rar K^\ast \ell^+ \ell^-$ and $B \rar K \ell^+ \ell^-$ decays. when
one uses the results of both Collaborations on these branching ratios,
stronger restrictions are imposed on some of the new Wilson coefficients. 
For example, $-2 \le C_{LL} \le 0$,
$0 \le C_{RL} \le 2.3$, $-1.5 \le C_T \le 1.5$ and $-3.3 \le C_{TE} \le
2.6$, and all of the remaining coefficients vary in the region $-4 \le C_X
\le 4$. The experimental results on the $B \rar K^\ast \ell^+ \ell^-$
and $B \rar K \ell^+ \ell^-$ decays are preliminary and for this reason we
did not take into account the above--mentioned restrictions. Moreover, we
assume that all new Wilson coefficients are real, as well as all of the form
factors that we use in the present work. As a result the normal polarization
of $\Lambda$ is equal to zero, since it is proportional to the imaginary
parts of the combinations of the new Wilson coefficients and of the form
factors.  

Before proceeding further with the numerical analysis, few words about lepton 
polarizations are in order. It follows from explicit expressions of the $\Lambda$
baryon polarizations that they depend on both $s$ and the new Wilson coefficients.
Therefore it may experimentally be difficult to study their dependence
on both of these variables simultaneously. For this reason it is better if we 
eliminate the dependence
of the $\Lambda$ baryon polarization on one of the variables. We choose to eliminate
the variable $s$ by performing integration over $s$ in the allowed
kinematical region, so that $\Lambda$ baryon polarizations are averaged over. The
averaged $\Lambda$ baryon polarizations are defined as
\bea
\label{e21}
\lla P_i \rra = \frac{\ds \int P_i \frac{d{\cal B}}{ds} ds} 
{\ds \int \frac{d{\cal B}}{ds} ds}~.
\eea

The dependence of the averaged lepton polarizations $\lla P_L\rra$ and
$\lla P_N\rra$ on the new Wilson coefficients are 
shown in Figs (1)--(2). From these figures we obtain the following results.

\begin{itemize}

\item $\lla P_L\rra$ is strongly dependent to the tensor interaction and
quite sensitive to the Wilson coefficients $C_{RR}$ and $C_{RL}$, for both 
$\Lambda_b \rar \Lambda \mu^+ \mu^-$ and $\Lambda_b \rar \Lambda \tau^+
\tau^-$ channels. We observe from Fig. (1) that the value of $\lla P_L \rra$
is negative for all values of the new Wilson coefficients
for $\Lambda_b \rar \Lambda \mu^+ \mu^-$ decay, while, as can easily be seen
from Fig. (2), it is positive when
$C_T \le -1.7$ and $C_{TE} \ge 0.5$ for the $\Lambda_b \rar \Lambda \tau^+
\tau^-$ channel. The $C_X=0$ point corresponds to the SM case. It follows from
Figs. (1) and (2) that the departure from the SM becomes substantial when
$C_X \neq 0$. This result confirms that the measurement of the longitudinal
$\Lambda$ baryon polarization can be a very decisive tool in looking for new
physics beyond the SM.   

\item The situation for $\lla P_N\rra$ is drastically different compared to
that for $\lla P_L\rra$. $\lla P_N\rra$ is strongly dependent to the vector
interaction coefficient $C_{RL}$ for both channels. The $\tau$ channel is
also sensitive to the tensor interaction coefficient $C_{TE}$ when $C_{TE} >
0$. It follows from Figs. (3) and (4) that $\lla P_N\rra$ is positive
(negative) when $C_{RL} < 0$ ($C_{RL} > 0$). In the $\tau$ channel $\lla
P_N\rra$ is negative when $C_{TE}$ positive. 

\end{itemize}

From these discussions we can conclude that change in sign and magnitude of
the $\Lambda$ baryon polarization is an indication of the new physics beyond
the SM. Determining the sign of the $\Lambda$ baryon polarization determines
the sign of the new Wilson coefficients and type of the new interactions. 

Finally we would like to mention about the branching ratio, whose
measurement is easier compared to that of the measurement of the 
$\Lambda$ baryon polarization, as well as being an efficient tool in 
establishing the new physics beyond the SM. In this connection, there follows
the question: can one establish new physics by concentrating on the lepton
polarization only. In other words, are there are certain
regions of the new Wilson coefficients in which the value of the branching
ratio coincides with that of the SM prediction, while
$\Lambda$ baryon polarization does not? In order to answer this we study the
correlation between the branching ratio and the averaged $\Lambda$ baryon
polarizations by varying the new Wilson coefficients. In further analysis
the values of the branching ratio ranges between $10^{-6} \le {\cal
B}(\Lambda_b \rar \Lambda \mu^+ \mu^-) \le 6 \times 10^{-6}$ and 
$10^{-7} \le {\cal B}(\Lambda_b \rar \Lambda \tau^+ \tau^-) \le 6 \times
10^{-7}$, which are of the same order of magnitude with the SM predictions.
A first glance to the analysis depicted in Figs. (5)--(9) yields the
following results.

\begin{itemize}

\item The numerical analysis for the $\Lambda_b \rar \Lambda \mu^+ \mu^-$ decay
for the longitudinal $\Lambda$ baryon polarization yields that such regions of 
$C_X$, in which branching ratio does agree with the SM prediction while 
the averaged longitudinal $\Lambda$ baryon polarization does not, are
absent. However, as can easily be seen  from Figs. (5) and (6), such regions
of $C_X$ exist $(C_{LR}$ and $C_{RL})$ for the averaged normal polarization 
of the $\Lambda$ baryon.

\item For the $\Lambda_b \rar \Lambda \tau^+ \tau^-$ decay, we observe the
existence of a very narrow region for the vector interaction coefficient
$C_{RL}$, in which the branching ratio coincides with the SM prediction but
the averaged longitudinal polarization does not. However, a study of the
correlation between the averaged normal polarization of the $\Lambda$
baryon and branching ratio, depicted in Figs. (7), (8) and (9), leads to 
more promising expectations. In other words, for the new Wilson coefficients 
$C_{LR},~C_{RLLR}$ and $C_{RLRL}$ there indeed exist such regions 
where branching ratio coincides with the SM prediction, but $\lla P_N\rra$ 
deviates substantially from that of the SM prediction.

\end{itemize}

In conclusion, we have studied the sensitivity of the $\Lambda$ baryon
polarizations to the new Wilson coefficients. It is shown that there exist
certain regions of various new Wilson coefficients for which the branching
ratio of the $\Lambda_b \rar \Lambda \ell^+ \ell^-$ decays coincide with the
SM prediction, while $\Lambda$ baryon polarizations deviate substantially 
from its counterparts predicted by the SM.

\newpage

\section*{Figure captions}
{\bf Fig. (1)} The dependence of the averaged longitudinal $\Lambda$ baryon
polarization $\lla P_L \rra$ on the new Wilson coefficients for the
$\Lambda_b \rar \Lambda \mu^+ \mu^-$ decay.\\ \\
{\bf Fig. (2)} The same as in Fig. (1), but for the 
$\Lambda_b \rar \Lambda \tau^+ \tau^-$ decay.\\ \\
{\bf Fig. (3)} The same as in Fig. (1), but for the averaged normal
$\Lambda$ baryon polarization $\lla P_N\rra$.\\ \\
{\bf Fig. (4)} The same as in Fig. (3), but for the            
$\Lambda_b \rar \Lambda \tau^+ \tau^-$ decay.\\ \\
{\bf Fig. (5)} Parametric plot of the correlation between the branching
ratio ${\cal B}$ and the averaged normal 
polarization $\lla P_N \rra$ as a function of the new vector $C_{LL}$ and
$C_{LR}$ Wilson coefficients
for the $\Lambda_b \rar \Lambda \mu^+ \mu^-$ decay.\\ \\
{\bf Fig. (6)} The same as in Fig. (5), but for the new vector $C_{RR}$ and
$C_{RL}$ Wilson coefficients.\\ \\
{\bf Fig. (7)} The same as in Fig. (5), but for the            
$\Lambda_b \rar \Lambda \tau^+ \tau^-$ decay.\\ \\
{\bf Fig. (8)} The same as in Fig. (6), but for the            
$\Lambda_b \rar \Lambda \tau^+ \tau^-$ decay.\\ \\
{\bf Fig. (9)} Parametric plot of the correlation between the branching
ratio ${\cal B}$ and the averaged normal 
polarization $\lla P_N \rra$ as a function of the new scalar
$C_{LRRL},~C_{LRLR},~C_{RLLR}$ and $C_{RLRL}$ and
$C_{LR}$ Wilson coefficients
for the $\Lambda_b \rar \Lambda \tau^+ \tau^-$ decay.\\ \\

\newpage

\begin{figure}
\vskip 1.5 cm
    \includegraphics{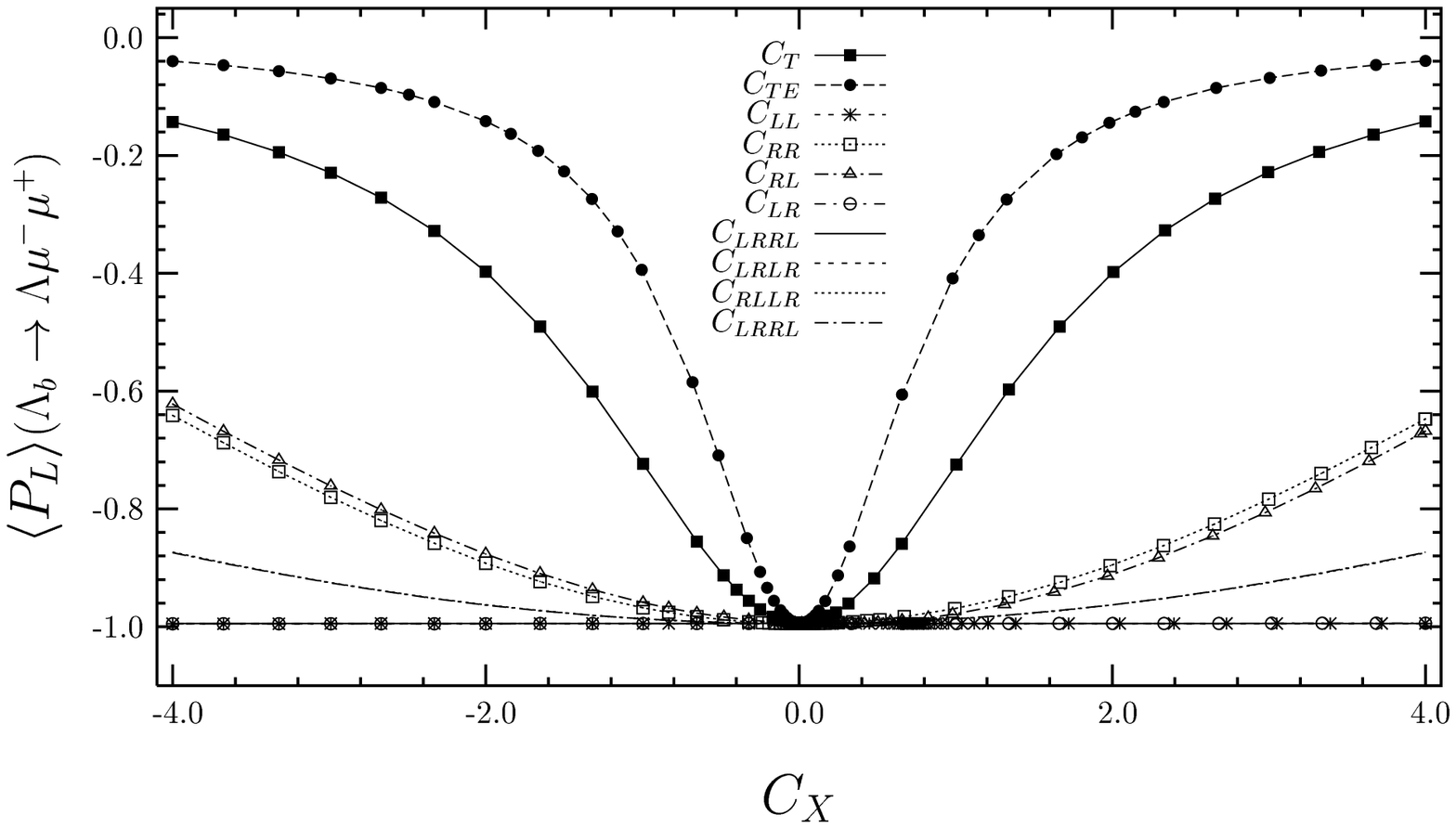}
\vskip 7.8cm
\caption{}
\end{figure}  

\begin{figure}   
\vskip 2.5 cm
    \includegraphics{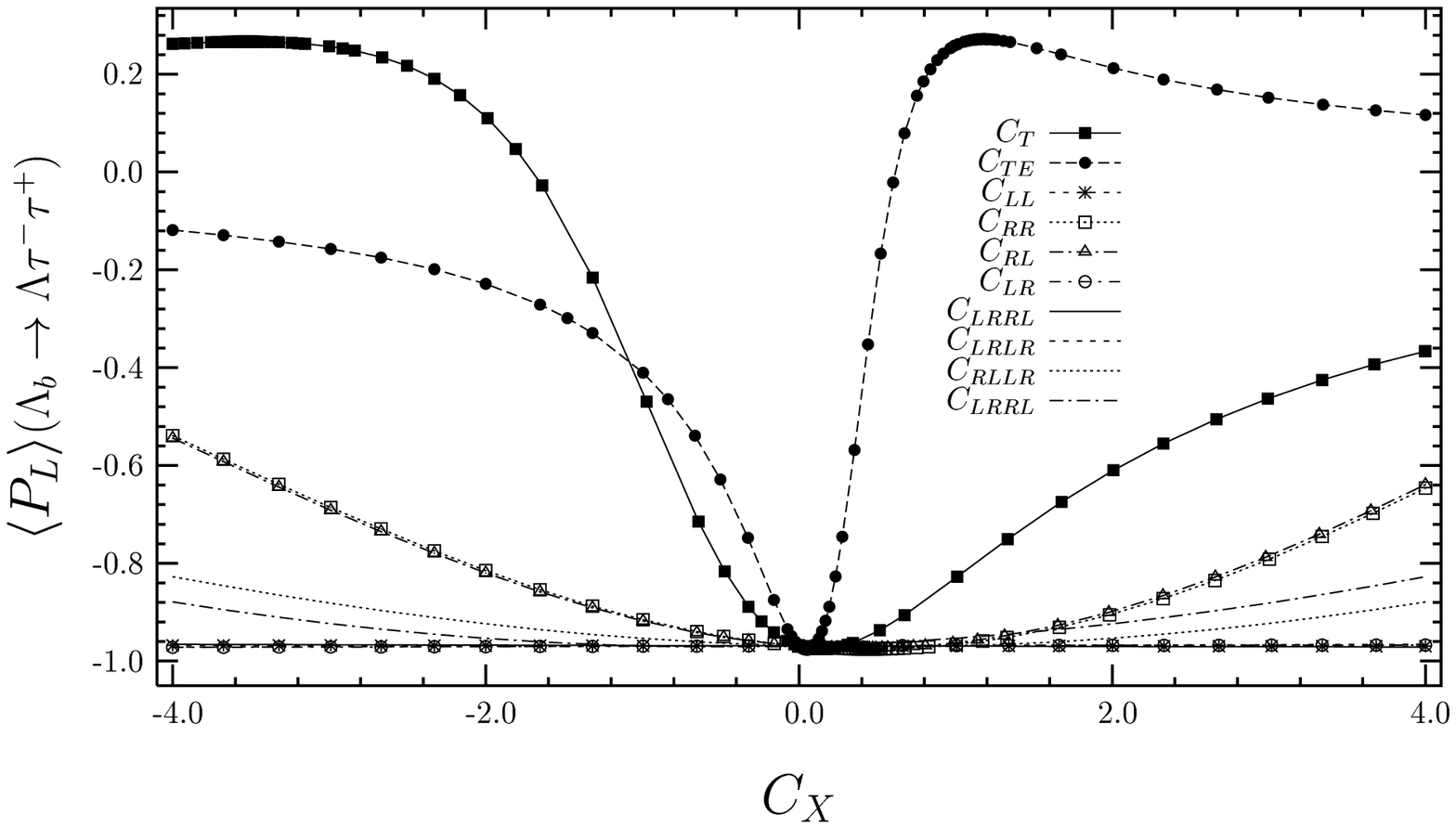}
\vskip 7.8 cm   
\caption{}
\end{figure}

\begin{figure}   
\vskip 1.5 cm
    \includegraphics{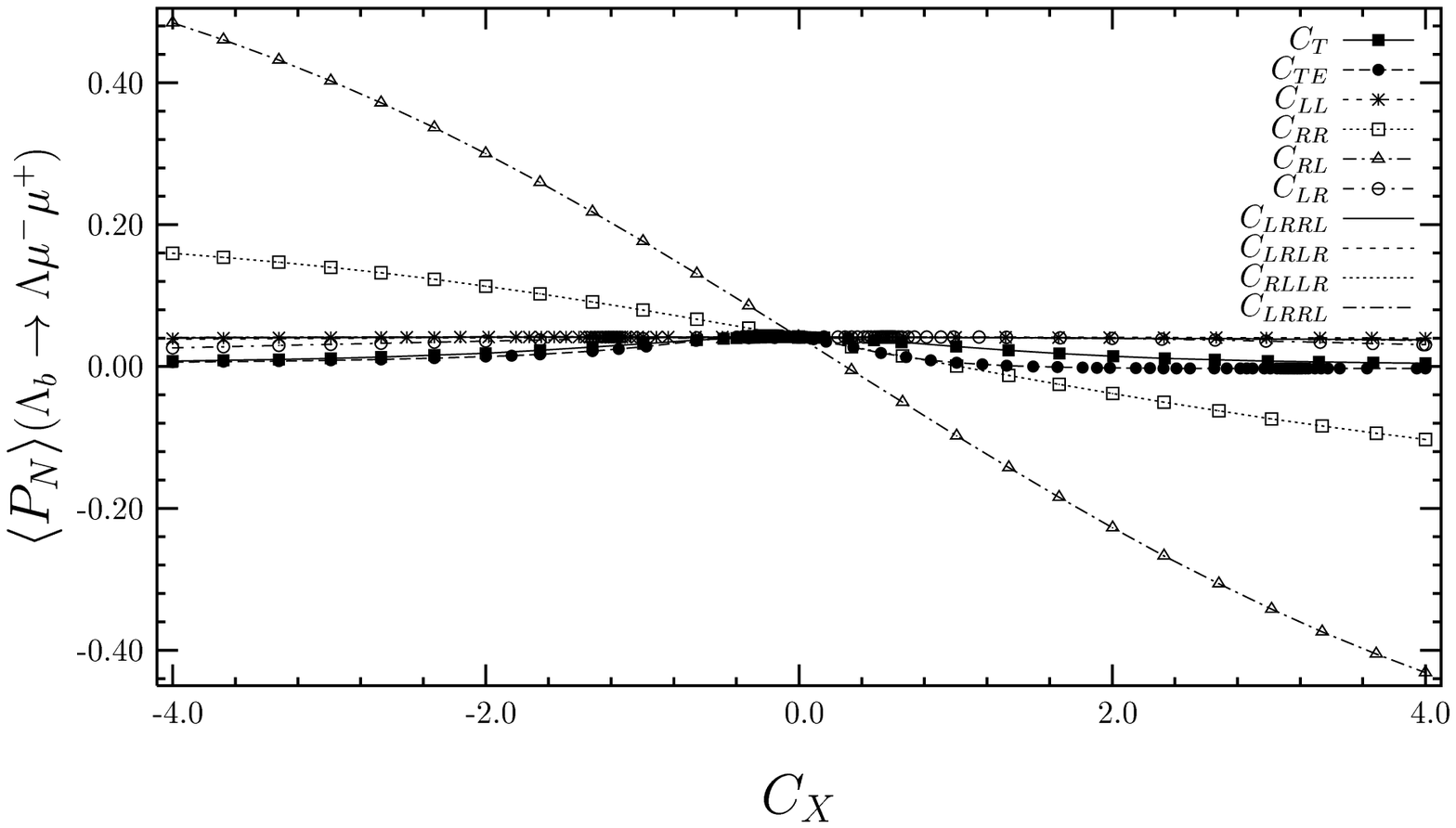}
\vskip 7.8cm
\caption{}
\end{figure}

\begin{figure}    
\vskip 2.5 cm
    \includegraphics{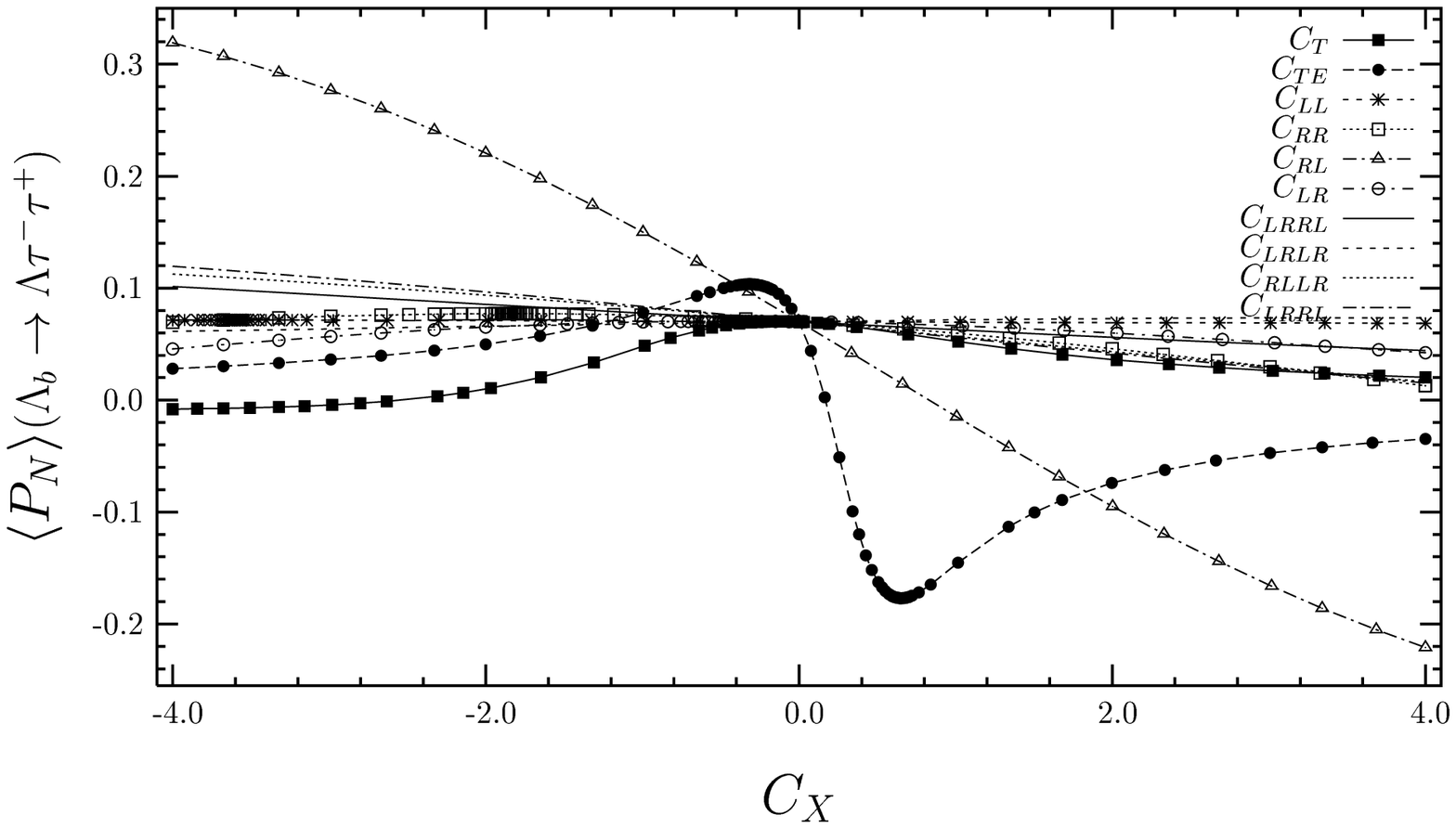}
\vskip 7.8 cm   
\caption{}
\end{figure}

\begin{figure}
\vskip 1.5 cm
    \includegraphics{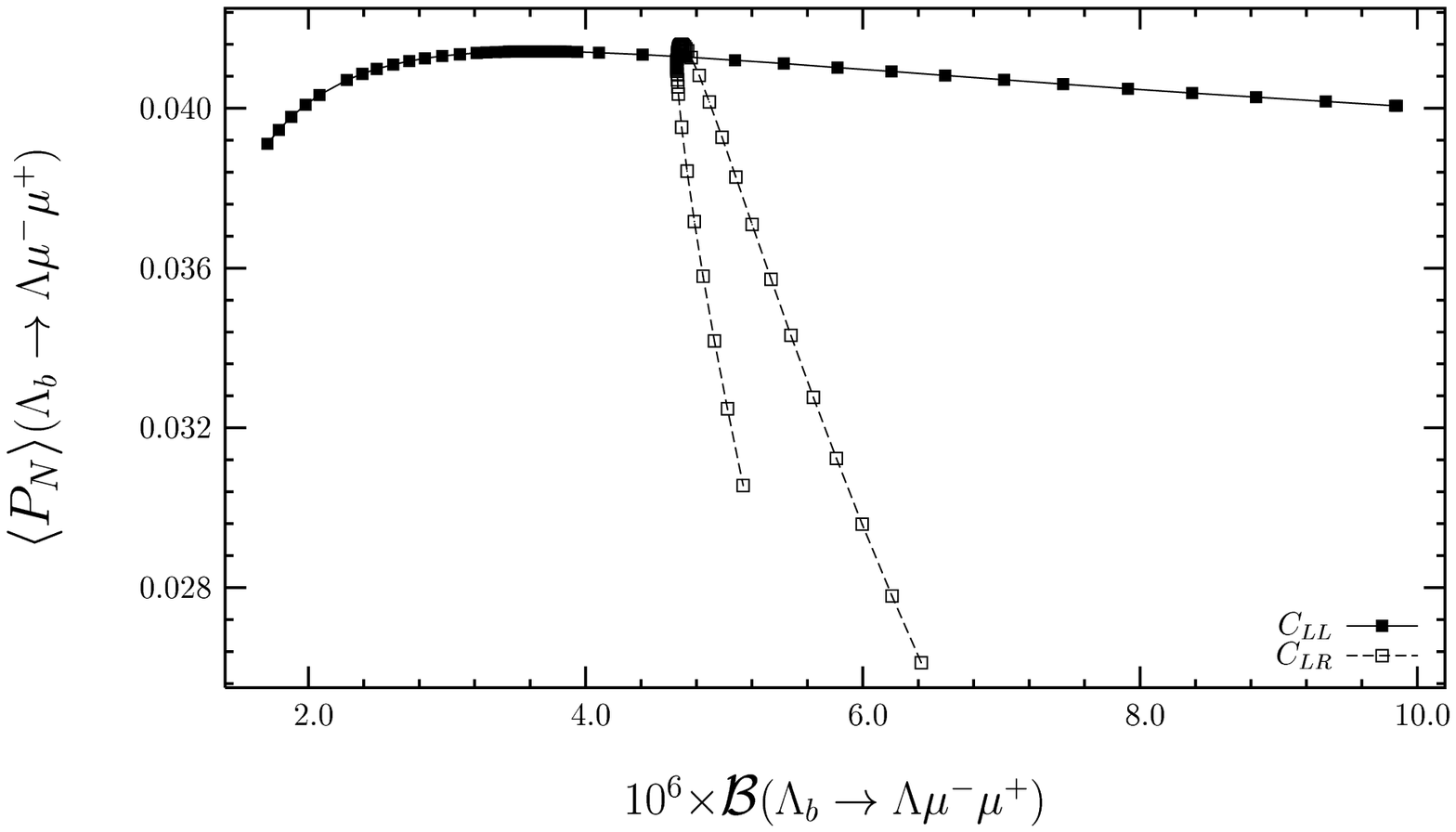}
\vskip 7.8cm
\caption{}
\end{figure}  

\begin{figure}   
\vskip 2.5 cm
    \includegraphics{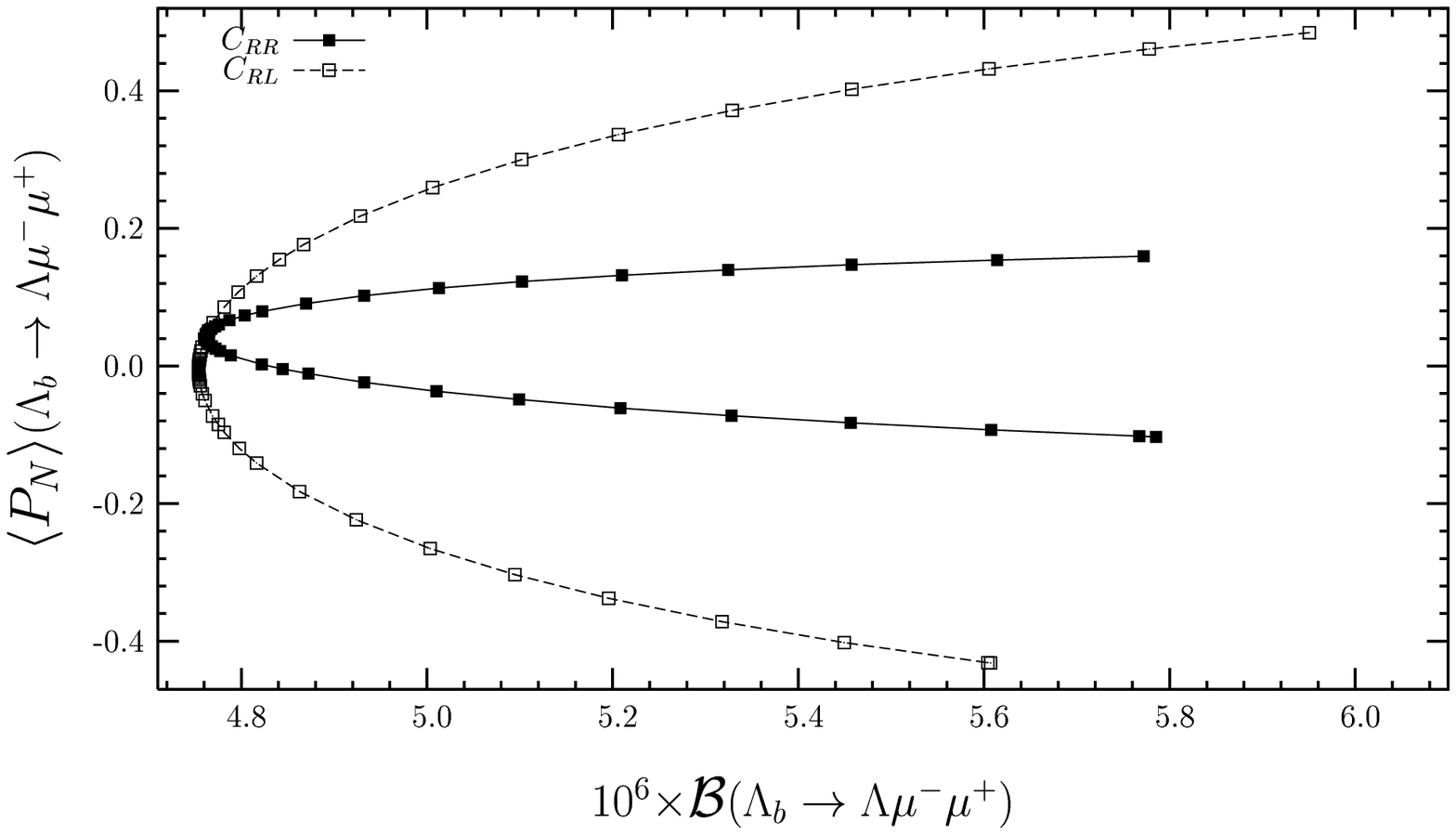}
\vskip 7.8 cm   
\caption{}
\end{figure}

\begin{figure}   
\vskip 1.5 cm
    \includegraphics{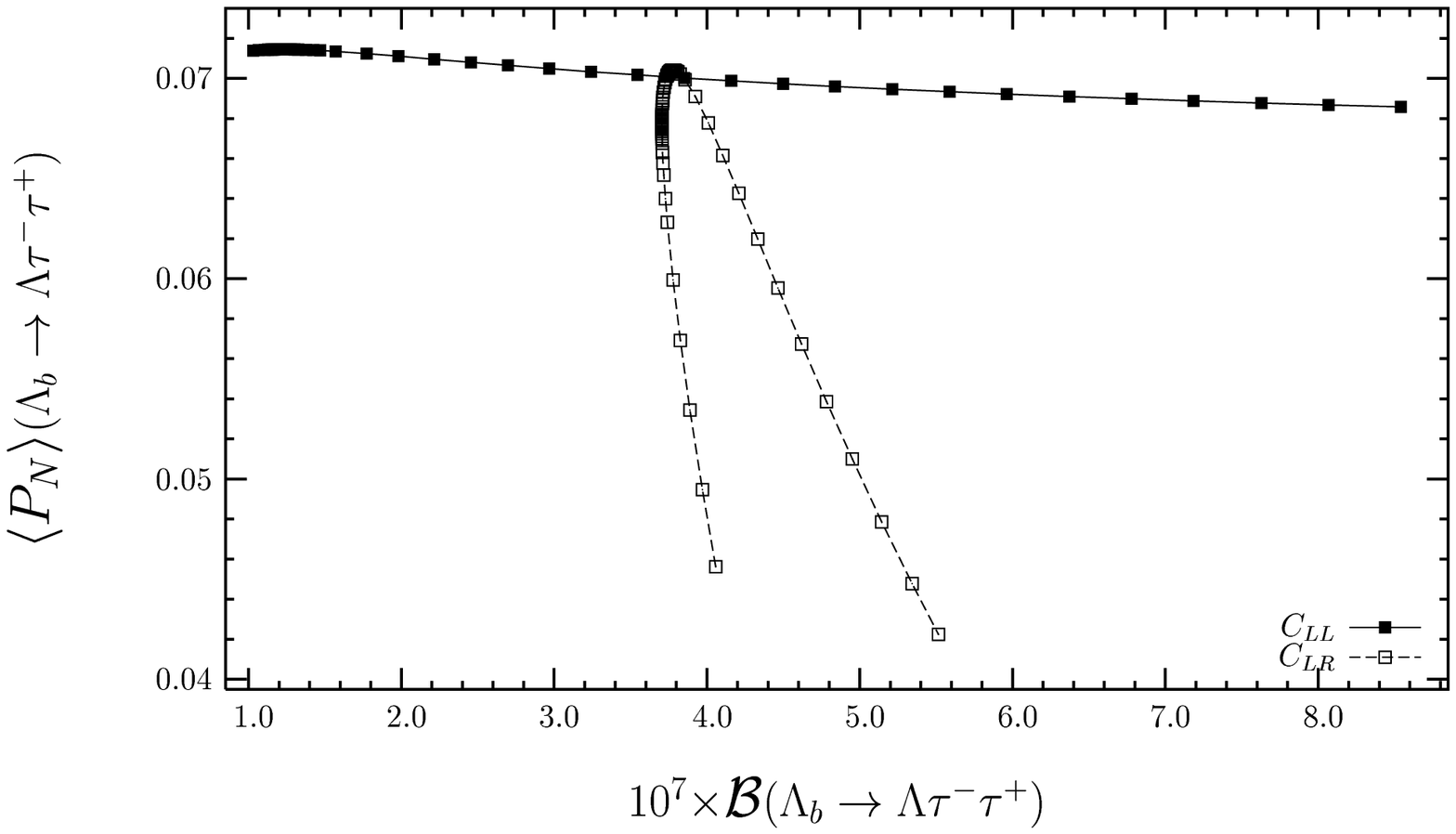}
\vskip 7.8cm
\caption{}
\end{figure}

\begin{figure}    
\vskip 2.5 cm
    \includegraphics{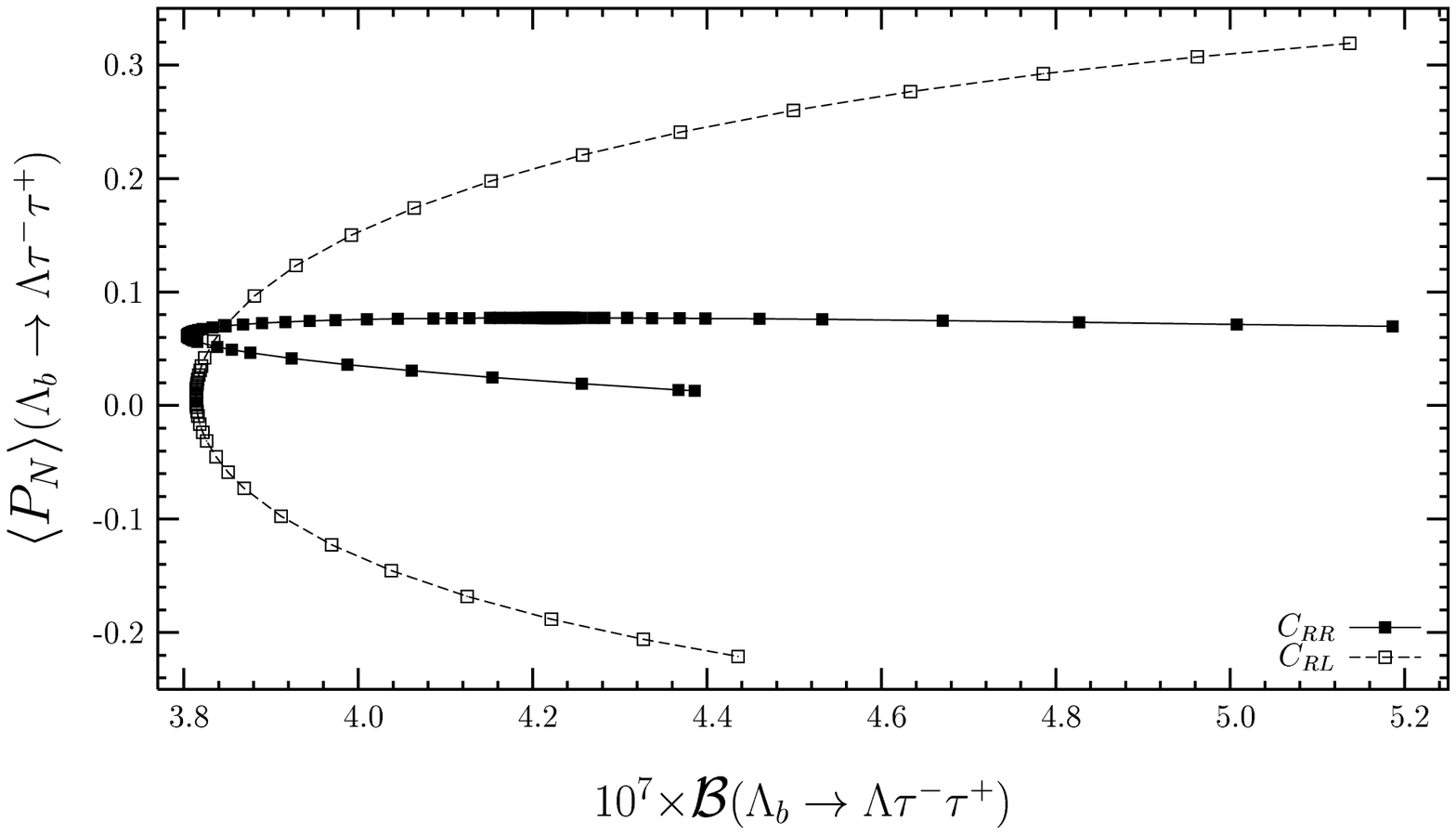}
\vskip 7.8 cm   
\caption{}
\end{figure}

\begin{figure}    
\vskip 2.5 cm
    \includegraphics{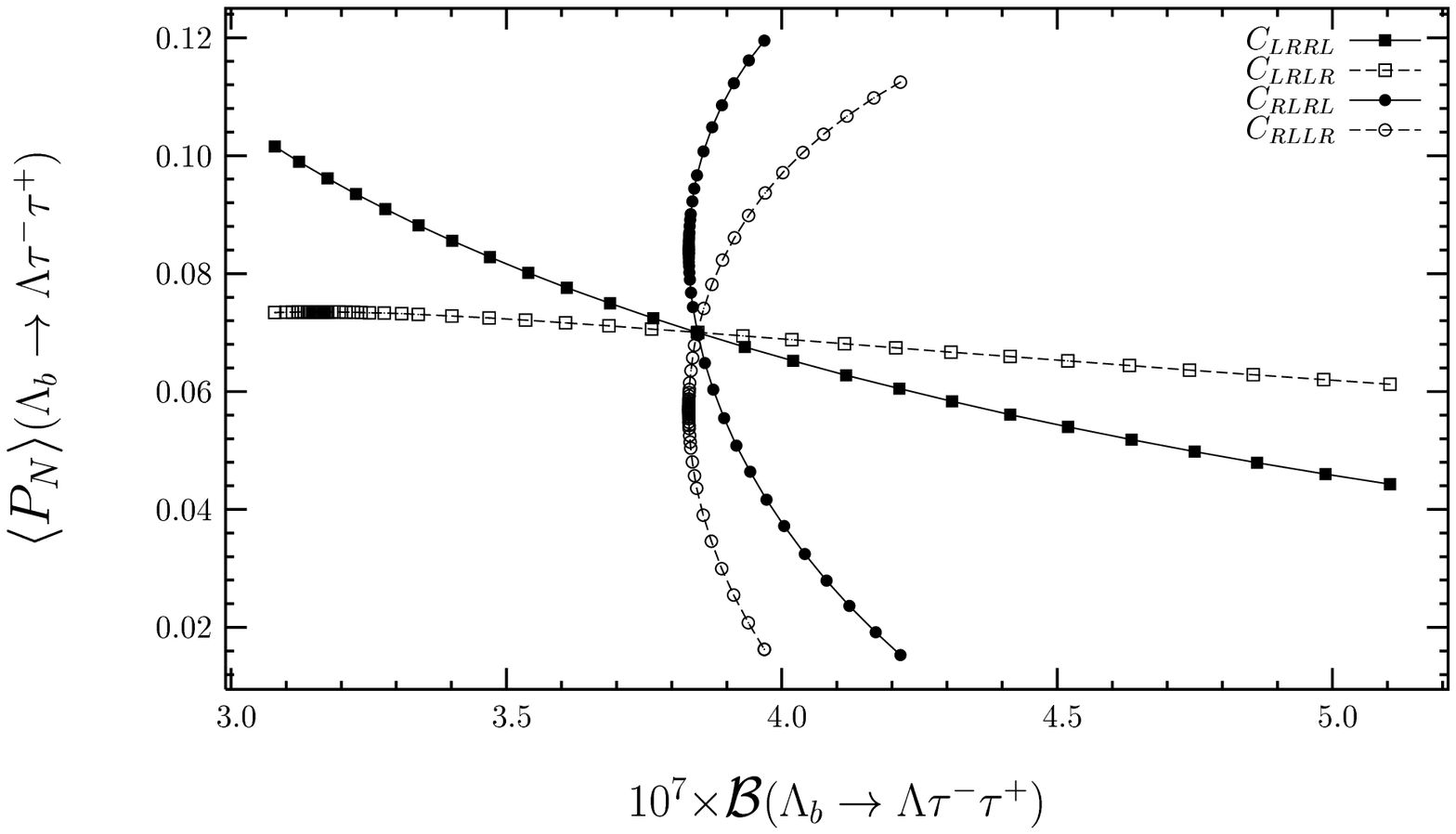}
\vskip 7.8 cm   
\caption{}
\end{figure}

\end{document}